\documentclass[10pt, twocolumn, comsoc]{IEEEtran}
\def\argmax{\mathop{\rm arg\,max}}

\usepackage{graphicx,epsfig}
\usepackage[noadjust]{cite}
\usepackage{mcite}
\usepackage{amsfonts,helvet}
\usepackage{fancyhdr}
\usepackage{threeparttable}
\usepackage{epsf,epsfig}
\usepackage{amsthm}
\usepackage{amsmath}
\usepackage{boldline}
\usepackage{booktabs}
\usepackage{siunitx}
\usepackage{amssymb}
\usepackage{lipsum}
\usepackage{epstopdf}

\usepackage{subcaption}
\usepackage{gensymb}

\usepackage{dsfont}
\usepackage{color}
\usepackage{array}
\usepackage{algpseudocode}
\usepackage{algcompatible}
\usepackage{enumerate}
\usepackage{cancel}

\usepackage[ruled]{algorithm2e}

\usepackage{siunitx}
\usepackage{fancyhdr}
\usepackage{wrapfig}
\usepackage{ragged2e}
\usepackage{algpseudocode}
\usepackage{algcompatible}
\usepackage{threeparttable}
\allowdisplaybreaks

\newtheorem{lemma}{Lemma}

\usepackage{eucal}

\begin{document}

\title{Space-Time Beamforming\\ for LEO Satellite Communications}

\author{Jungbin Yim,~\IEEEmembership{Student Member,~IEEE,} Jinseok Choi,~\IEEEmembership{Member,~IEEE,} Jeonghun Park,~\IEEEmembership{Member,~IEEE,} \\Ian P. Roberts,~\IEEEmembership{Member,~IEEE,} and Namyoon Lee,~\IEEEmembership{Senior Member,~IEEE} 



\thanks{J. Yim is with Department of Electrical Engineering, POSTECH, Pohang, South Korea (email:{\texttt{jungbinyim@postech.ac.kr}}), J. Choi is with School of Electrical Engineering, KAIST, Daejeon, South Korea (email:{\texttt{jinseok@kaist.ac.kr}}), J. Park is with School of Electrical and Electronic Engineering, Yonsei University, Seoul, South Korea (e-mail:{\texttt{jhpark@yonsei.ac.kr}}), I. P. Roberts is with Department of Electrical and Computer Engineering, UCLA, Los Angeles, CA 90095, USA (e-mail:{\texttt{ianroberts@ucla.edu}}), N. Lee is with Department of Electrical Engineering, POSTECH, Pohang, South Korea (e-mail:{\texttt{nylee@postech.ac.kr}}).  
}}

\maketitle

\begin{abstract}
Inter-beam interference poses a significant challenge in low Earth orbit (LEO) satellite communications due to dense satellite constellations. To address this issue, we introduce space-time beamforming, a novel paradigm that leverages the space-time channel vector, uniquely determined by the angle of arrival (AoA) and relative Doppler shift, to optimize beamforming between a moving satellite transmitter and a ground station user. We propose two space-time beamforming techniques: space-time zero-forcing (ST-ZF) and space-time signal-to-leakage-plus-noise ratio (ST-SLNR) maximization. In a partially connected interference channel, ST-ZF achieves a 3dB SNR gain over the conventional interference avoidance method using maximum ratio transmission beamforming. Moreover, in general interference networks, ST-SLNR beamforming significantly enhances sum spectral efficiency compared to conventional interference management approaches. These results demonstrate the effectiveness of space-time beamforming in improving spectral efficiency and interference mitigation for next-generation LEO satellite networks.

\end{abstract}


\section{Introduction}
Satellite communication in low Earth orbit (LEO) can offer high-speed Internet connectivity with low delays on a global scale. These satellite networks complement traditional terrestrial networks, especially in areas where coverage is limited \cite{cioni2018satellite,guidotti2019architectures,lin2021path,Lin_Lee_book}. Consequently, LEO satellite networks have emerged as the cornerstone of next-generation wireless communication systems, enabling global connectivity and bridging the digital divide \cite{LEO-mega}. The deployment of dense satellite constellations with overlapping beam coverages has significantly expanded communication capacity and reduced latency. However, this dense configuration also introduces a critical challenge: beam interference \cite{LEO-SG,LEO-IM, LEO-BP,TWC2023,TWC2024}.


Beam interference occurs when multiple satellites operate concurrently, often serving co-located or geographically proximate users, leading to overlapping beams and significant mutual interference \cite{LEO-IM, LEO-BP}. This problem is exacerbated by the limited aperture size of satellite antennas, dynamic relative motion of satellites, and aggressive frequency reuse strategies employed to maximize spectral efficiency. Such interference reduces spectral efficiency and limits the overall network capacity, posing a bottleneck for LEO satellite systems \cite{LEO-SG,TWC2023,TWC2024}.

Traditional spatial beamforming techniques have been considered as an effective solution to mitigate the interference problem. However, as the density increases and antenna aperture sizes are limited, they offer limited flexibility in managing interference, particularly when multiple satellites attempt to serve co-located users simultaneously \cite{LEO-BP,fortes2003impact,tonkin2018newspace,braun2019should}. Another approach is to allocate different frequency resources across adjacent satellites to avoid the possible interference effect such as band-splitting \cite{pai2017update,braun2019should}. However, the rapid motion of LEO satellites introduces time-varying interference patterns that complicate the allocation of frequencies dynamically based on real-time interference conditions. 


In this paper, we tackle the challenge of beam interference in LEO satellite networks by introducing space-time beamforming, a novel approach inspired by space-time adaptive processing (STAP) \cite{ward1995space,STAP1} and synthetic-aperture radar (SAR) processing \cite{sar}. Unlike existing space-time transmission techniques\textemdash such as space-time coding \cite{tarokh1998space}, space-time interference alignment \cite{stia,dstia,Lee_Heath_2016}, and space-time network coding \cite{stnc}\textemdash the proposed method leverages STAP's ability to jointly exploit spatial and temporal diversity for signal enhancement in dynamic environments. Originally developed for radar systems, STAP has a potential in LEO satellite communication systems where both spatial and temporal channel characteristics fluctuate. Motivated by its success in the radar domain, we extend STAP to multi-antenna LEO satellite communications as an effective means of interference mitigation.


\subsection{Contribution}
 We summarize our contributions of this paper.
 
 \begin{itemize}
 	\item  We first formulate the space-time beamforming design problem, a novel interference mitigation paradigm that leverages the space-time channel vector, defined by the angle of arrival (AoA) and relative Doppler shift, to optimize beamforming in time-varying LEO satellite channels. In this approach, a moving satellite transmitter repeatedly transmits a data symbol at specific intervals while aligning the beam direction with the intended ground user's AoA and normalized Doppler frequency. This space-time beamforming strategy effectively mitigates co-channel interference among spatially co-located users by exploiting their distinct Doppler characteristics. However, the repetition of transmissions introduces a fundamental trade-off: while it enhances interference suppression, it also reduces spectral efficiency. To maximize sum spectral efficiency in interfering LEO satellite networks, space-time beamforming must be jointly optimized across three key parameters: (i) the number of repetitions, (ii) the repetition intervals, and (iii) the space-time beam direction. 
 	\item  We propose two space-time beamforming techniques that optimize key design parameters in two interference network scenarios: (i) partially connected and (ii) fully connected interference network models. In a partially connected interference network, where a ground user experiences strong interference from an adjacent satellite transmitter, we introduce space-time zero-forcing (ST-ZF) beamforming, which jointly optimizes (i) the number of repetitions, (ii) the repetition intervals, and (iii) the space-time beam direction.  A key finding of our study is that two repetitions provide the optimal balance between interference suppression and spectral efficiency, and the optimal retransmission interval should be chosen such that space-time channels become orthogonal. This structural optimization ensures effective interference mitigation while maintaining transmission efficiency. With this optimal design, we demonstrate that ST-ZF achieves a 3 dB SNR gain over the conventional maximum ratio transmission (MRT) with time-division multiple access (TDMA), which was previously considered information-theoretically optimal for partially connected interference networks when the channel remains constant over time \cite{etkin2008gaussian,jafar2010generalized}. 

\item Next, we generalize space-time beamforming to a fully connected $K$-user interference network by introducing space-time signal-to-leakage-plus-noise ratio (ST-SLNR) beamforming. To optimize the key design parameters, we develop a suboptimal space-time beamforming technique that sequentially determines these parameters in a structured manner. Leveraging the eigenvector structure of SLNR beamforming, we first optimize the space-time beam direction for a given repetition interval and number of repetitions. Then, using a one-dimensional grid search, we determine the optimal repetition interval that maximizes the SLNR. Finally, we optimize the number of repetitions to maximize sum spectral efficiency. Additionally, we extend this beamforming algorithm to scenarios with imperfect channel knowledge at the satellite, ensuring robustness in practical LEO satellite networks where precise channel state information (CSI) may not always be available. This generalization highlights the adaptability of space-time beamforming in mitigating interference under realistic channel conditions.
\item Simulation results confirm that the proposed ST-SLNR beamforming achieves significant sum spectral efficiency gains compared to traditional interference mitigation techniques, including conventional SLNR beamforming and TDMA with spatial beamforming. These results demonstrate the effectiveness of space-time beamforming in enhancing interference suppression and optimizing spectral efficiency in fully connected LEO satellite interference networks. This result highlights the effectiveness of space-time beamforming in improving interference management and spectral efficiency in LEO satellite networks.



 \end{itemize}

%

 \subsection{Organization}
Section II provides an overview of the spatial-temporal channel model. In Section III, we formulate the space-time beamforming problem. Section IV presents the ST-ZF beamforming technique for partially connected interference networks, while Section V introduces ST-SLNR beamforming for general interference networks. Simulation results are discussed in Section VI, and conclusions are drawn in Section VII.

\section{System Model} \label{sec:system model}
In this section, we describe the network model along with the assumed CSI knowledge and present the channel model.

\subsection{Network Model }
We consider a downlink LEO satellite communication system where the satellite is equipped with a uniform planar array (UPA) consisting of $N_x$ antennas along the $x$-axis and $N_y$ antennas along the $y$-axis. Consequently, the total number of antennas in the UPA is $N=N_x\times N_y$. Furthermore, we assume that the satellites have CSI at the transmitter (CSIT) for the links between themselves and the associated users. This assumption can be justified by the fact that users are capable of estimating their own valid positions using systems such as global navigation satellite system (GNSS), and subsequently report this location information to the satellites via the radio access network (RAN) \cite{3gpp-ts-38.300,3gpp-tr-38.811,3gpp-tr-38.821}. Based on the received location information, satellites can access the users’ positions and thereby determine the direction and relative velocity required for providing user-specific services under LOS channel scenarios. In non-LOS channel scenarios, each satellite may acquire the CSIT from the uplink pilots sent by the ground users.  
 
 We consider two interference network models: i) partially-connected and ii) fully-connected interference networks.  A partially connected interference network arises when each user experiences strong interference from only a subset of transmitters, while others contribute negligible interference due to path loss, beam misalignment, or frequency separation. In LEO satellite systems, this occurs when adjacent satellites cover overlapping regions, creating localized interference clusters. This structure allows for efficient interference mitigation, such as ST-ZF beamforming, which will be exaplined in Section IV.

A fully connected interference network occurs when each user receives interference from all $K-1$ transmitters. This scenario is common in multi-beam LEO satellite systems, where beams are overlapped from multiple satellites. Managing interference in such networks requires ST-SLNR beamforming, which will be explained in Section \uppercase\expandafter{\romannumeral4}.
 

\subsection{Channel Model}
We define the spatial array response at time \( t \) for satellite \( k \), which is located at position \( \mathbf{p}_k^t \in \mathbb{R}^3 \), with respect to user \( \ell \). Let \( \theta_{\ell,k,i}^t \) and \( \phi_{\ell,k,i}^t \) represent the zenith and azimuth angles of the \( i \)th propagation path between satellite \( k \) and user \( \ell \) at time \( t \), measured relative to the broadside of the uniform planar array (UPA).  The corresponding spatial array response vector at time \( t \) is then given by:
 \begin{align}
{\bf a} &\left(\theta_{\ell,k,i}^t,\phi_{\ell,k,i}^t\right)=\nonumber\\
&\mathbf{\bar a}_x\left(\sin\theta_{\ell,k,i}^t\cos\phi_{\ell,k,i}^t\right)\otimes\mathbf{\bar a}_y\left(\sin\theta_{\ell,k,i}^t\sin\phi_{\ell,k,i}^t\right)\in \mathbb{C}^{N\times 1}, \label{eq:spatial steering vector}
\end{align}
with 
\begin{align}
    \mathbf{\bar a}_x(u)=\begin{bmatrix}
        1 & e^{-j\frac{2\pi}{\lambda}d u} & \cdots & e^{-j\frac{2\pi}{\lambda}(N_x-1)d u}
    \end{bmatrix}^{\top}\in \mathbb{C}^{N_x\times 1},  
\end{align} and
\begin{align}
    \mathbf{\bar a}_y(u)=\begin{bmatrix}
        1 & e^{-j\frac{2\pi}{\lambda}d u} & \cdots & e^{-j\frac{2\pi}{\lambda}(N_y-1)d u}
    \end{bmatrix}^{\top}\in \mathbb{C}^{N_y\times 1},  
\end{align}
where $\lambda$ and $d$ are the carrier wavelength and inter-antenna spacing. In addition, $\otimes$ denotes the Kronecker product of two vectors.
 
Using the spatial array response vector, we define the downlink channel for the link between satellite \( k \) and user \( \ell \) at an initial time \( t \), denoted as \( {\bf h}_{\ell,k}[1] \). This channel is modeled as a superposition of multiple propagation paths as  
\begin{align}
    {\bf h}_{\ell,k}[1] = \sum_{i=1}^{L_{\ell,k}}\beta_{\ell,k,i}^t {\bf a}  \left(\theta_{\ell,k,i}^t, \phi_{\ell,k,i}^t\right)\in \mathbb{C}^{N\times 1}, \label{eq:spatial channel model}
\end{align}  
where \( L_{\ell,k} \) denotes the number of propagation paths in the channel between satellite \( k \) and user \( \ell \), and \(\beta_{\ell,k,i}^t \) represents the attenuation constant for the \( i \)th path at time \( t \). This attenuation factor incorporates the transmission power of satellite \( k \), the path-loss determined by the distance between satellite \( k \) and user \( \ell \), and the channel fading power, which accounts for shadowing and small-scale fading effects.

We define the space-time channel as a combination of spatial and temporal steering vectors. To construct this channel model, we first describe how the spatial steering vector is defined at time slot \( t+(m-1)\tau_k \) for \( m \in [M] \). Specifically, at time \( t+(m-1)\tau_k \), satellite \( k \) moves to position \( \mathbf{p}_{k}^{t+(m-1) \tau_k} \) with a relative velocity \( v_{\ell,k} \) with respect to user \( \ell \).  

Following the principles of STAP in \cite{ward1995space,STAP1,sar}, we assume that the time duration over \( M \) samples, \( (M-1)\tau_k \), is sufficiently small such that the AoAs and channel coefficients remain time-invariant. This implies  
\[
\theta_{\ell,k,i}^{t+(m-1)\tau_k} \simeq \theta_{\ell,k,i}^{t}, \quad \phi_{\ell,k,i}^{t+(m-1)\tau_k} \simeq \phi_{\ell,k,i}^{t},  
\]
and
\[
\beta_{\ell,k,i}^{t+(m-1)\tau_k} \simeq \beta_{\ell,k,i}^t,
\]
for $m\in [M]$ and $ i \in [L_{\ell,k}]$. From this point, for ease of exposition, we omit the time index for AoAs and the channel coefficients, i.e.,  $\theta_{\ell,k,i}^{t}=\theta_{\ell,k,i}$, $\phi_{\ell,k,i}^{t}=\phi_{\ell,k,i}$, and $\beta_{\ell,k,i}^t=\beta_{\ell,k,i}$. Under this time-invariant AoAs and channel coefficients  assumption during the co-processing interval, the spatial steering vector at time \( t+(m-1)\tau_k \), with the satellite located at \( \mathbf{p}_{k}^{t+(m-1) \tau_k} \), can be expressed using the initial channel vector \( \mathbf{h}_{\ell,k}[1] \) and its Doppler shift \( f_{\ell,k} = v_{\ell,k} / \lambda \) as  
\begin{align}
  \mathbf{h}_{\ell,k}[m]  = \mathbf{h}_{\ell,k}[1]e^{-j2\pi(m-1)f_{\ell,k}\tau_k} \in \mathbb{C}^{M\times 1}. \label{eq:spatial channel model2}
\end{align}  
By stacking the Doppler shifts across time intervals, we define the temporal steering vector, which captures the Doppler effect, as  
\begin{equation}
    \mathbf{b}(f_{\ell,k},\tau_k) \!=\!
    \begin{bmatrix}
        1\! & e^{-j2\pi f_{\ell,k}\tau_k}\! &\! \cdots &\! e^{-j2\pi(M-1) f_{\ell,k}\tau_k}
    \end{bmatrix}^{\!\top} \!\!\!\in \!\mathbb{C}^{M\times 1}. \label{eq:temporal steering vector}
\end{equation}  
The space-time channel vector over the time interval \( (M-1)\tau_k \) is then obtained using the weighted sum of the Kronecker product between the spatial steering vector \( \mathbf{a}(\theta_{\ell,k,i},\phi_{\ell,k,i}) \) in \eqref{eq:spatial steering vector} and the temporal steering vector \( \mathbf{b}(f_{\ell,k},\tau_k) \) in \eqref{eq:temporal steering vector}, as  
\begin{align}
   \mathbf{h}_{\ell,k}^{M,\tau_k}= \sum_{i=1}^{L_{\ell,k}} \beta_{\ell,k,i} {\bf c}_{\ell,k,i}^{M,\tau_k}  \in \mathbb{C}^{MN\times 1}, \label{eq:STchannel}
\end{align}
where
\begin{align}
   {\bf c}_{\ell,k,i}^{M,\tau_K}  =   \mathbf{b}(f_{\ell,k},\tau_k) \otimes \mathbf{a}(\theta_{\ell,k,i},\phi_{\ell,k,i})  \in \mathbb{C}^{MN\times 1}. \label{eq:space-time steering vector}
\end{align}
This space-time channel is uniquely determined by the angle of arrival \( (\theta_{\ell,k,i}, \phi_{\ell,k,i}) \), the satellite-induced Doppler shift \( f_{\ell,k} \), and the temporal sampling interval \( \tau_k \), which together define its structure.

\subsection{Virtual Array Interpretation}
\begin{figure}[t!]
  \centering
  \includegraphics[width=0.47\textwidth]{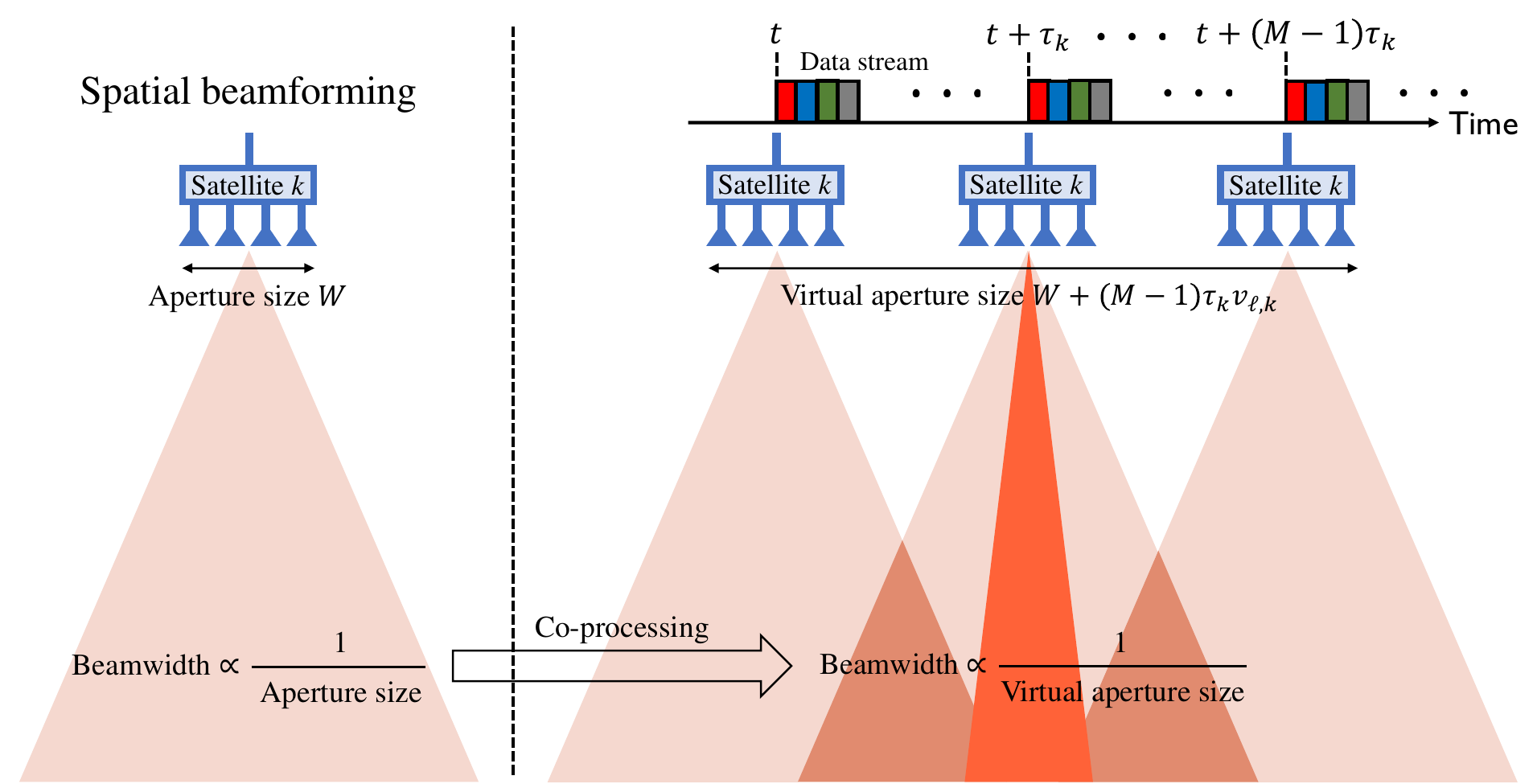}
  \caption{Example illustrating the concept of virtual array interpretation.}
  \label{fig:virtual array interpretation} \vspace{-0.2cm}
\end{figure}

The space-time channel vector in \eqref{eq:STchannel} can be interpreted as a superposition of virtual array response vectors, similar to those found in STAP and SAR systems. To illustrate this connection, we consider a special case where only the line-of-sight (LOS) path exists, i.e., \( L_{\ell,k}=1 \). In this scenario, dropping the path index \( i \), the space-time channel simplifies to  
\begin{align}
	 \mathbf{h}_{\ell,k}^{M,\tau_k}=\beta_{\ell,k} \mathbf{c}_{\ell,k}^{M,\tau_k} \in \mathbb{C}^{MN\times 1}. \label{eq:LOSchannel}
\end{align}  
This space-time channel vector can be interpreted as a virtual array response as shown in Fig. \ref{fig:virtual array interpretation}, where the virtual array takes the form of an array-of-subarrays with an inter-subarray spacing of \( \tau_{k} v_{\ell,k} \). Given that the satellite's physical antenna array has an aperture size of \( W \), the effective size of the virtual array expands to \( W + (M-1) \tau_{k} v_{\ell,k} \). This virtual array synthesis extends the effective aperture beyond the satellite's physical constraints, allowing for improving spatial resolution to separate users via narrow beams because the beam width is inversely proportional to the aperture size of the array.   Additionally, the temporal channel structure, shaped by Doppler frequencies and signal repetition patterns, introduces an additional distinguishing feature. This provides a unique signature that helps differentiate co-located users who cannot be resolved using spatial resolution alone, enhancing interference mitigation and improving spectral efficiency.


\subsection{Space-Time Transmission and Reception} 
To construct the space-time channel from satellite \( k \) to user \( \ell \), denoted as \( {\bf h}_{\ell,k}^{M,\tau_k} \) in \eqref{eq:STchannel}, satellite \( k \) transmits a data symbol \( s_k \in \mathbb{C} \) repeatedly \( M \) times at uniform time intervals \( \tau_k \). The time interval \( \tau_k \) is assumed to be an integer multiple of the sample period \( T_s = \frac{1}{B} \), where \( B \) is the signal bandwidth, i.e.,  
\begin{align}
	\tau_k = rT_s,
\end{align}  
where \( r \in \mathbb{Z}^{+} \) is a positive integer.  

 Satellite \( k \) transmits the data symbol \( s_k \sim \mathcal{CN}(0,1) \) using a space-time beamforming vector with transmit power $P$. Therefore, we simply consider the data symbol as \( s_k \sim \mathcal{CN}(0,P) \). Let \( {\bf f}_k[m] \in \mathbb{C}^{N\times 1} \) be the transmit beamforming vector carrying the information symbol \( s_k \) at time slot \( m \in [M] \). The received signal at user \( k \) at time slot \( m \) is given by  
\begin{equation}
    y_{k}[m] = {\bf h}_{k,k}^{\sf H}[m]\mathbf{f}_{k}[m]s_k + \sum_{\ell \neq k}^{K} {\bf h}_{k,\ell}^{\sf H}[m]{\bf f}_{\ell}[m]s_{\ell} + n_{k}[m], \label{eq:mth received signal}
\end{equation}  
where \( n_{k}[m] \) is the additive white Gaussian noise at user \( k \) at time slot \( m \), distributed as \( \mathcal{CN}(0,\sigma^2) \).  

Defining the stacked space-time beamforming vector carrying the data symbol \( s_k \) over \( M \) time slots as  
\begin{align}
	{\bf f}_k^M = \begin{bmatrix}
        {\bf f}_k^{\top}[1] & {\bf f}_k^{\top}[2] & \cdots & {\bf f}_k^{\top}[M]
    \end{bmatrix}^{\top} \in \mathbb{C}^{MN\times 1}, 
\end{align}  
we similarly denote the stacked space-time channel vector from satellite \( k \) to user \( \ell \), parameterized by \( (M,\tau_k) \), as  
\begin{align}
	{\bf h}_{k,\ell}^{M, \tau_{\ell}}= \begin{bmatrix}
        {\bf h}_{k,\ell}^{\top}[1] & {\bf h}_{k,\ell}^{\top}[2] & \cdots & {\bf h}_{k,\ell}^{\top}[M]
    \end{bmatrix}^{\top} \in \mathbb{C}^{MN\times 1}. 
\end{align}  
Applying a simple receive processing that combines the \( M \) received signals, the total received signal is given by  
\begin{align}
  y_{k}^M = \left({\bf h}_{k,k}^{M,\tau_k}\right)^{\sf H}{\bf f}_{k}^Ms_k + \sum_{\ell \neq k}^{K} \left({\bf h}_{k,\ell}^{M,\tau_{\ell}}\right)^{\sf H}{\bf f}_{\ell}^M s_{\ell} + n_{k}^M, \label{eq:received signal}
\end{align}  
where \( y_{k}^M = \sum_{m=1}^{M} y_{k}[m] \) and \( n_{k}^M = \sum_{m=1}^{M} n_{k}[m] \sim \mathcal{CN}(0,M\sigma^2) \).  From \eqref{eq:received signal}, the signal-to-interference-plus-noise ratio (SINR) at user $k$ is defined as  
\begin{align}
	{\sf SINR}_{k}\left(M,\left\{\tau_k , {\bf f}_k^M\right\}_{k=1}^K\right) = \frac{\left|\left({\bf h}_{k,k}^{M,\tau_k}\right)^{\sf H}{\bf f}_k^M\right|^2 P}{\sum_{\ell\neq k}^K \left|\left({\bf h}_{k,\ell}^{M,\tau_{\ell}}\right)^{\sf H}{\bf f}_{\ell}^M\right|^2 P + M\sigma^2},
\end{align}  
where \( \|{\bf f}_k^M\|_2^2 = M \) for \( k \in [K] \). The achievable spectral efficiency at user \( \ell \) is then given by  
\begin{align}
R_{k}\left(M,\left\{\tau_k , {\bf f}_k^M\right\}_{k=1}^K\right) \!=\!\frac{1}{M} \log_2 \left( 1\! +\! {\sf SINR}_{k}\left(M,\left\{\tau_k , {\bf f}_k^M\right\}_{k=1}^K\right) \right), \label{eq:space-time sum rate}
\end{align}  
where the pre-log term \( 1/M \) accounts for the rate penalty due to repeated transmissions. To avoid notational ambiguity in the algorithm pseudocode presented in Section \uppercase\expandafter{\romannumeral4}, we define the achievable spectral efficiency for $M=0$ and $M=1$ as $R_{k}\left(0\right)$ and $R_{k}\left(1,\left\{{\bf f}_k^1\right\}_{k=1}^K\right)$, omitting parameters that are not required.

Our goal is to optimize the space-time beamforming vector \( \{{\bf f}_k^M\} \), the number of repeated transmissions \( M \), and the time interval \( \{\tau_k\} \) to maximize the achievable sum spectral efficiency under the local CSIT assumption. The optimization problem is formulated as  
\begin{equation}
    \begin{aligned}
       \max_{M, \{\tau_k\}, \{{\bf f}_k^M\}} \quad & \sum_{k=1}^K R_{k}\left(M,\left\{\tau_k , {\bf f}_k^M\right\}_{k=1}^K\right) \\ 
        \text{s.t.} \quad & \|{\bf f}_k^M\|_2^2 = M, \quad k \in [K].
    \end{aligned}
    \label{eq:optimization}
\end{equation}  
 For a given \( \tau_k \) and \( M \), finding the optimal beamforming vectors \( {\bf f}_k^M \) for \( k \in [K] \) is a well-known NP-hard problem. Existing approaches such as weighted minimum mean squared error (WMMSE) beamforming \cite{WMMSE}, fractional programming \cite{FP}, and generalized power iteration precoding (GPIP) \cite{GPIP} can provide suboptimal solutions under global CSIT, which requires CSI to be shared among satellites. In LEO satellite systems, sharing CSIT across satellites introduces significant signaling overhead, making centralized beamforming approaches impractical. As a result, beamforming techniques based on local CSIT as in \cite{GPIP_localCSIT} are more suitable for real-world deployment.  


\section{Space-Time Zero-Forcing Beamforming} \label{sec:partially connected interference channel}
\begin{figure}[t!]
  \centering
\includegraphics[width=0.47\textwidth]{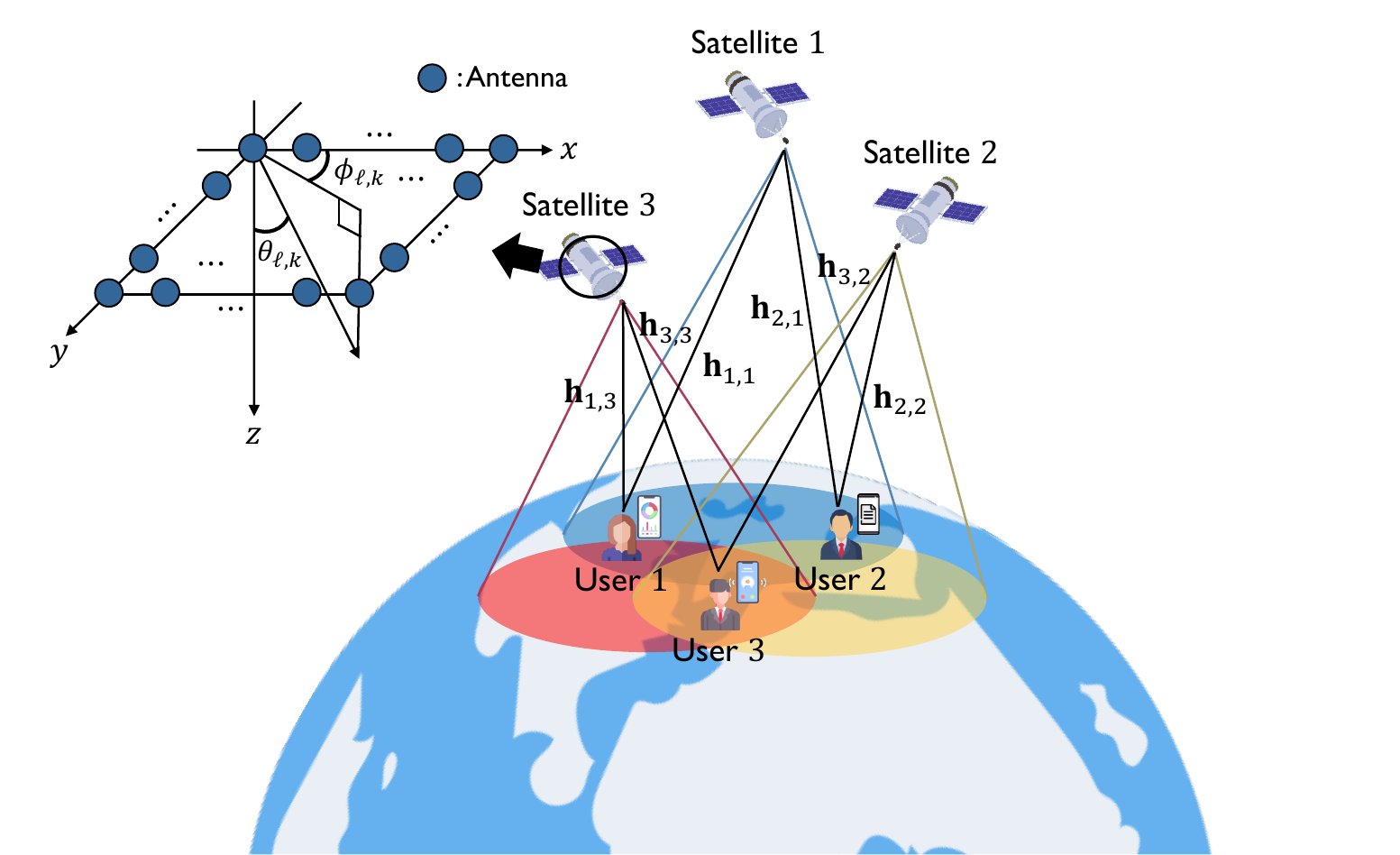}
  \caption{Illustration of the downlink LEO satellite communication system under a partially connected interference channel ($K=3$).}
  \label{fig:Network model in special case} \vspace{-0.2cm}
\end{figure}

To illustrate the principle of space-time beamforming, we introduce a simple yet effective technique called ST-ZF beamforming in this section. Specifically, we focus on a partially connected \( K \)-user interference network as shown in Fig. \ref{fig:Network model in special case}, where satellite $k$ serves user $k$ while interfering adjacent user $k'$ with $k' = k+1$ for $k\in [K-1]$ and $k'=1$ for $k=K$.  In addition, we consider a line-of-sight (LOS)-dominant channel model, where the channel vectors are formed solely by the LOS path. Furthermore, we assume that the AoAs from satellite $k$ to the desired user $k$ and interfering user $k'$ are identical, i.e., $\theta_{k,k}=\theta_{k',k}$ and $\phi_{k,k}=\phi_{k',k}$, capturing the scenario where  user \( k \) and user $k'$ are located within the beam coverage area of both satellite \( k \).

\subsection{Traditional Approach}
Under this setup, we first provide a brief review of the traditional approach where space-time beamforming is not applied. This serves as a baseline for understanding the impact and benefits of the proposed space-time beamforming technique. 
%

When no space-time beamforming is applied, i.e., with no time extension (\( M=1 \)),  the channel vectors from satellite \( k \) and the adjacent interfering satellite \(k''\) to user $k$ are given by
\[
{\bf h}_{k,k}[1] = \beta_{k,k} {\bf a}(\theta_{k,k},\phi_{k,k})
\]
and  
\[
{\bf h}_{k, k''}[1] = \beta_{k, k''} {\bf a}(\theta_{k, k''},\phi_{k, k''}),
\]
where $k''=k-1$ for $k\in \{2,\ldots, K\}$ and $k''=1$ for $k=K$. In this case, user \( k \) receives the data symbol \( s_k \) transmitted from satellite \( k \) along with beamforming vector ${\bf f}_k[1]$ as  
\begin{align}
	y_k[1] =  {\bf h}_{k,k}^{\sf H}[1]{\bf f}_k[1]s_k + {\bf h}_{k,k''}^{\sf H}[1] {\bf f}_{k''}[1]s_{k''} +n_k[1].
\end{align} 
Since the AoAs are assumed to be identical for both the desired and interfering links, the two channel vectors become aligned, meaning they span the same subspace, i.e.,
\begin{align}
	{\rm rank}\left( \begin{bmatrix}
        {\bf h}_{k,k}[1] &{\bf h}_{k,k''}[1]
    \end{bmatrix}\right)=1.
\end{align}
As a result, conventional spatial beamforming alone cannot effectively mitigate interference, leading to significant performance degradation. In this scenario, the standard approach for mitigating interference is TDMA. Assuming \( K \) is an even number, the satellites are divided into two distinct groups: \( \mathcal{K}_{\text{odd}} = \{1, 3, \ldots, K-1\} \) and \( \mathcal{K}_{\text{even}} = \{2, 4, \ldots, K\} \).  To prevent interference, during odd time slots, satellites in \( \mathcal{K}_{\text{odd}} \) transmit data symbols using maximum ratio transmission (MRT) beamforming, given by  
\[
\mathbf{f}_{k}[1] = \frac{\mathbf{h}_{k,k}[1]}{\|\mathbf{h}_{k,k}[1]\|_2}, \quad k \in \mathcal{K}_{\text{odd}},
\]  
while satellites in \( \mathcal{K}_{\text{even}} \) remain silent. In even time slots, the roles are reversed, with satellites in \( \mathcal{K}_{\text{even}} \) transmitting while those in \( \mathcal{K}_{\text{odd}} \) remain silent.  This interference avoidance scheme is considered information-theoretically optimal from a degrees-of-freedom (DoF) perspective, providing a capacity approximation in high signal-to-noise ratio (SNR) conditions \cite{TIM}. The achievable sum spectral efficiency for this interference management technique becomes
\begin{align}
	R_{\sf sum}^{\sf TDMA}=\frac{1}{2}\sum_{k=1}^K\log_2\left(1+\frac{|\beta_{k,k}|^2 NP}{\sigma^2}\right). \label{eq:SR_1}
\end{align}

\subsection{Space-Time Beamforming}
 We will demonstrate that the proposed space-time beamforming technique enhances sum spectral efficiency by effectively leveraging spatial-temporal channel properties, particularly when satellites move at different relative velocities with respect to the users.  

Consider a scenario where satellite \( k \) moves with velocity \( v_k \) and transmits the information symbol \( s_k \) twice with a time interval of \( \tau_k \), corresponding to \( M=2 \). The satellite employs precoding vectors \( {\bf f}_k[1] \) and \( {\bf f}_k[2] \) for each transmission. The received signals at user \( k \) over two time slots are given by  
\begin{align}
    y_{k}[1] = {\bf h}_{k,k}^{\sf H}[1]\mathbf{f}_{k}[1]s_k + {\bf h}_{k,k''}^{\sf H}[1]{\bf f}_{k''}[1]s_{k''} + n_{k}[1], 
\end{align}  
where the interference channel is defined as  
\[
{\bf h}_{k,k''}^{\sf H}[1]=\beta_{k,k''}  \mathbf{a}(\theta_{k,k''},\phi_{k,k''}),
\]  
and for the second time slot, the received signal is  
\begin{align}
    y_{k}[2] = {\bf h}_{k,k}^{\sf H}[2]\mathbf{f}_{k}[2]s_k + {\bf h}_{k,k''}^{\sf H}[2]{\bf f}_{k''}[2]s_{k''} + n_{k}[2], 
\end{align}  
with  
\[
{\bf h}_{k,k''}^{\sf H}[2]=\beta_{k,k''}e^{-j2\pi f_{k,k''}\tau_{k''}}  \mathbf{a}(\theta_{k,k''},\phi_{k,k''}).
\]  
User \( k \) then combines the two received signals, resulting in the following space-time representation:  
\begin{align}
	y_k^2=\left({\bf h}_{k,k}^{2,\tau_{k}}\right)^{\sf H}{\bf f}_k^2 s_k + \left({\bf h}_{k,k''}^{2,\tau_{k''}}\right)^{\sf H}{\bf f}_{k''}^2 s_{k''} + n_{k}^2,
\end{align}  
where  
$y_k^2=\sum_{m=1}^2y_{k}[m]$,  $n_k^2=\sum_{m=1}^2n_{k}[m]\sim\mathcal{CN}(0,2\sigma^2),$ $
{\bf f}_k^2=\left[{\bf f}_k^{\top}[1],~{\bf f}_k^{\top}[2]\right]^{\top},$
and the space-time channel from satellite \( k \) to user \( k \) is given by  
\begin{align}
	{\bf h}_{k,k}^{2,\tau_{k}}=\beta_{k,k}  \left[1,~ e^{-j2\pi f_{k,k}\tau_k}\right]^{\top}\otimes \mathbf{a}(\theta_{k,k},\phi_{k,k}).
\end{align}  
Since satellite \( k \) has local CSIT, it can design the space-time beamforming vector \( {\bf f}_{k} \) using knowledge of \( {\bf h}_{k,k}^{2,\tau_{k}} \) and \( {\bf h}_{k',k}^{2,\tau_{k}} \). Although the AoAs from satellite \( k \) to its intended user \( k \) and interfering user \( k' \) are identical, the space-time channel vectors can be linearly independent, satisfying
\begin{align}
	{\rm rank}\left( \begin{bmatrix}
        {\bf h}_{k,k}^{2,\tau_{k}} &{\bf h}_{k',k}^{2,\tau_{k}}
    \end{bmatrix}\right)=2, 
\end{align} 
provided that the normalized Doppler frequencies satisfy  
\begin{align}
 	f_{k,k} \neq f_{k',k}+\frac{2\pi w}{\tau_k},
\end{align}  
where $w\in Z^+$ is an any integer. To verify this condition, we express the matrix consisting of the two space-time channel vectors as  
\begin{align}
 	&\begin{bmatrix}
        {\bf h}_{k,k}^{2,\tau_{k}} &{\bf h}_{k',k}^{2,\tau_{k}}
    \end{bmatrix}  = \nonumber \\ 
    &\qquad\beta_{k,k}  	\begin{bmatrix}
        1& 1 \\
            e^{-j2\pi f_{k,k}\tau_k}& e^{-j2\pi f_{k',k}\tau_k} 
    \end{bmatrix} \otimes \mathbf{a}(\theta_{k,k},\phi_{k,k}).
\end{align}  
Using the rank property of the Kronecker product,  
\begin{align}
	\text{rank}(\mathbf{A} \otimes \mathbf{B}) = \text{rank}(\mathbf{A}) \text{rank}(\mathbf{B}), \label{eq:rankprop}
\end{align}  
for any matrices \( \mathbf{A} \in \mathbb{C}^{m \times n} \) and \( \mathbf{B} \in \mathbb{C}^{p \times q} \), we conclude that \( {\bf h}_{k,k}^{2,\tau_{k}} \) and \( {\bf h}_{k',k}^{2,\tau_{k}} \) are linearly independent as long as  
\[
f_{k,k}\tau_{k} \neq f_{k',k}\tau_{k}+2\pi w.
\]  
In practice, this condition occurs when user \( k \) and user \( k' \) are not perfectly aligned with the satellite's flight direction, leading to distinct relative Doppler frequencies for the two users. This independence allows space-time beamforming to effectively separate signals, enhancing interference mitigation and improving spectral efficiency.

 Leveraging this linear independence, the proposed space-time beamforming aims to eliminate inter-satellite interference by ensuring that  
\begin{align}
	\left({\bf h}_{k',k}^{2,\tau_{k}}\right)^{\sf H}{\bf f}_{k}=0
\end{align}  
for all \( k \in [K] \). This constraint ensures that the interference from adjacent satellites is effectively suppressed.  We call this as ST-ZF beamforming.  With this ST-ZF, the achievable sum spectral efficiency is given by 
\begin{align}
	R_{\sf sum}^{\sf STZF}=\frac{1}{2}\sum_{k=1}^K\log_2\left(1+\frac{|\beta_{k,k}|^2 \left|\left({\bf c}_{k,k}^{2,\tau_{k}}\right)^{\sf H}{\bf f}_{k}^2\right|^2P}{2\sigma^2}\right). \label{eq:SR_2}
\end{align}  

Although ST-ZF beamforming successfully eliminates interference to adjacent cell users, the per-user effective SNR may be lower than that of conventional TDMA with spatial beamforming, as given in \eqref{eq:SR_1}. This reduction is due to the noise boosting effect caused by adding the noise signals from retransmission, expressed as  
\begin{align}
	 \frac{|\beta_{k,k}|^2\left|\left({\bf c}_{k,k}^{2,\tau_{k}}\right)^{\sf H}{\bf f}_{k}^2\right|^2P}{2\sigma^2} < \frac{|\beta_{k,k}|^2 NP}{\sigma^2}.
\end{align}  
However, space-time beamforming provides an additional degree of freedom through the choice of the retransmission interval \( \tau_k \). By carefully selecting \( \tau_k \), it is possible to create a favorable space-time channel to maximize the received signal power while satisfying   $\left({\bf h}_{k',k}^{2,\tau_{k}}\right)^{\sf H}{\bf f}_{k}=0$ as
\begin{align}
 	\arg\max_{\tau_k , {\bf f}_k^2 }\left|\left({\bf h}_{k,k}^{2,\tau_{k}}\right)^{\sf H}{\bf f}_{k}^2\right|^2.
 \end{align}  
The following lemma shows how to choose the retransmission interval $\tau_k$ to create a favorable space-time channel to maximize the space-time beamforming gain. 

\begin{lemma}
The two space-time channel vectors, \( {\bf h}_{k,k}^{2,\tau_{k}} \) and \( {\bf h}_{k',k}^{2,\tau_{k}} \) for $k\in [K]$, can be designed to be orthogonal, ensuring  
\begin{align}
 	\left({\bf h}_{k,k}^{2,\tau_{k}}\right)^{\sf H}{\bf h}_{k',k}^{2,\tau_{k}}=0,
\end{align}  
when the retransmission interval \( \tau_k \) is chosen as  
\begin{equation}
        \tau_k=\frac{1}{2(f_{k,k}-f_{k',k})}. \label{eq:retransmission_interval_of_ST-ZF}
\end{equation}  
\end{lemma}
This lemma implies that by appropriately selecting \( \tau_k \), space-time beamforming can effectively exploit Doppler-based diversity, ensuring interference-free transmission and improving system performance.

 \begin{proof}
Consider two vectors \( {\bf u}_i \in \mathbb{C}^{M\times 1} \) and \( {\bf v}_i \in \mathbb{C}^{N\times 1} \), where \( i \in \{1,2\} \). The inner product of their Kronecker products is given by  
\begin{align}
	({\bf u}_1\otimes {\bf v}_1)^{\sf H}({\bf u}_2\otimes {\bf v}_2) = ({\bf u}_1^{\sf H}{\bf u}_2) \cdot ({\bf v}_1^{\sf H}{\bf v}_2).
\end{align}  
Using this property, the inner product of the two space-time channel vectors can be computed as  
\begin{align}
	\left({\bf h}_{k,k}^{2,\tau_{k}}\right)^{\sf H}{\bf h}_{k',k}^{2,\tau_{k}} 
	&= \beta_{k,k} \beta_{k',k} {\bf b}^{\sf H}(f_{k,k},\tau_k){\bf b} (f_{k',k},\tau_k) \nonumber\\
	&~~~\times  \mathbf{a}^{\sf H}(\theta_{k,k},\phi_{k,k})  \mathbf{a}(\theta_{k,k},\phi_{k,k})\nonumber\\
	&= \beta_{k,k} \beta_{k',k} {\bf b}^{\sf H}(f_{k,k},\tau_k){\bf b} (f_{k',k},\tau_k) N.
\end{align}  
The inner product of the two temporal steering vectors is given by  
\begin{align}
	{\bf b}^{\sf H}(f_{k,k},\tau_k){\bf b} (f_{k',k},\tau_k) = \frac{\sin\left(2\pi\tau_k(f_{k,k}-f_{k',k})\right)}{\sin\left(\pi\tau_k(f_{k,k}-f_{k',k})\right)}.
\end{align}  
To ensure orthogonality, this inner product must be zero, which is satisfied when  
\begin{align}
 \tau_k^{\star} = \frac{1}{2(f_{k,k}-f_{k',k})}.
\end{align}  
This completes the proof.  
\end{proof}

\begin{figure}[t!]
  \centering
  \includegraphics[width=0.47\textwidth]{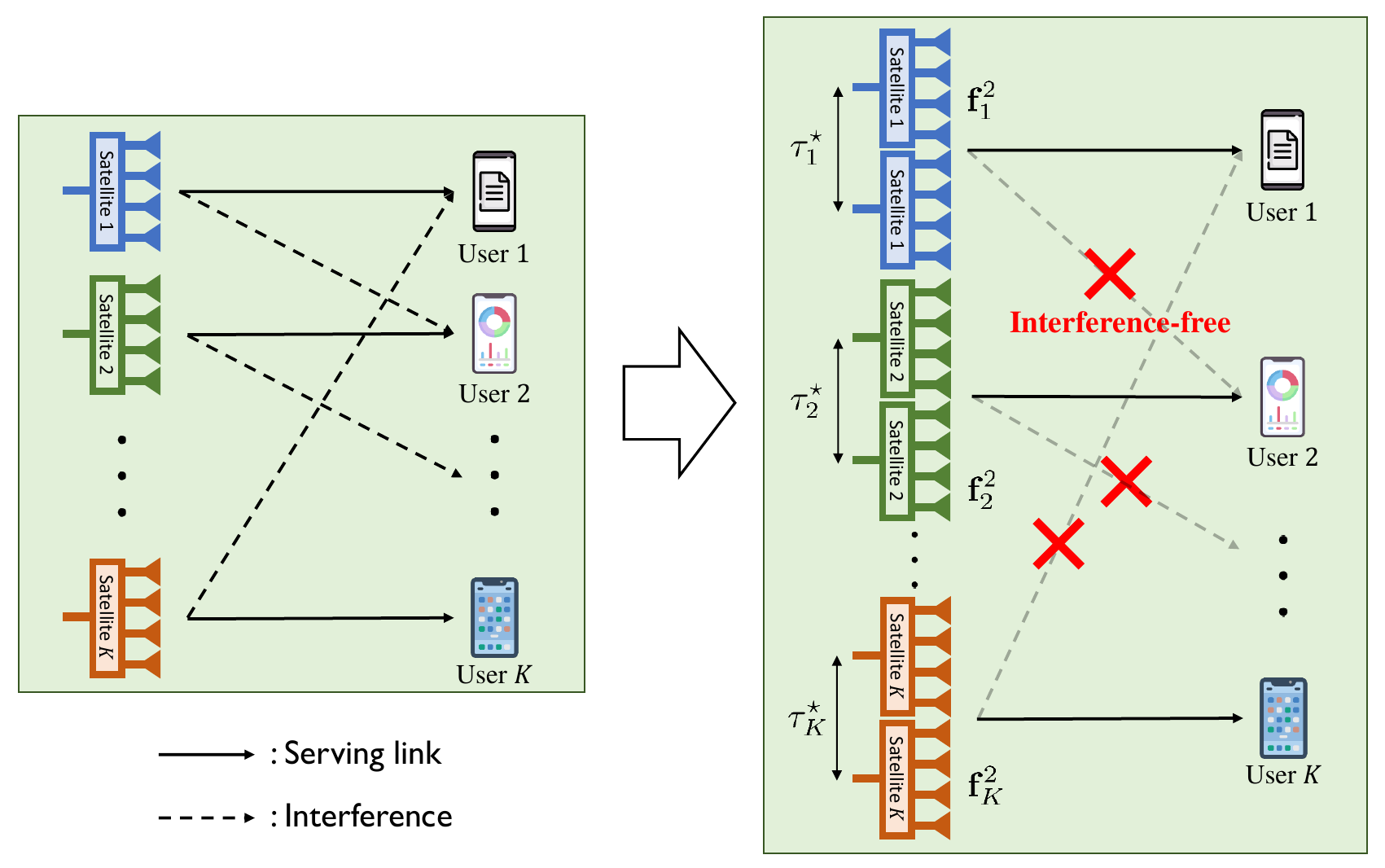}
  \caption{An illustration of the proposed ST-ZF method. The solid lines represent serving links, while the dashed lines indicate interference links.}
  \label{fig:STZF diagram} \vspace{-0.2cm}
\end{figure}

\begin{figure*}[t!]
  \centering
  \includegraphics[width=1.0\textwidth]{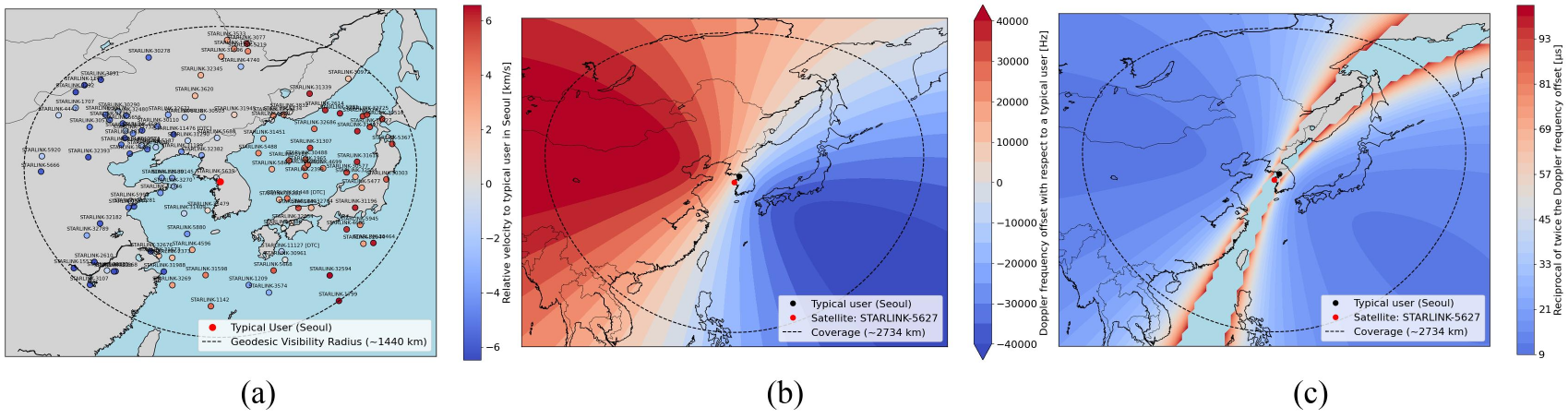}
  \caption{ Illustration of relative satellite motion and associated Doppler characteristics with respect to a typical user located in Seoul. (a) Relative velocity of visible satellites to the typical user. (b) Doppler frequency offset distribution observed by the selected satellite. (c) Reciprocal of twice the Doppler frequency offset, corresponding to the retransmission interval defined in \eqref{eq:retransmission_interval_of_ST-ZF}.  }
  \label{fig:Satellite_profiles} \vspace{-0.2cm}
\end{figure*}

From Lemma 1, it is possible to ensure that the desired and interfering channel vectors are orthogonal by setting the retransmission interval as  $\tau_k^{\star} = \frac{1}{2(f_{k,k}-f_{k',k})}$, which guarantees  $
\left({\bf h}_{k,k}^{2,\tau_{k}^{\star}}\right)^{\sf H}{\bf h}_{k',k}^{2,\tau_{k}^{\star}}=0.$
In this case, the MRT beamforming vector $\mathbf{f}_{k}^2 = \frac{\mathbf{h}_{k,k}^{2, \tau_k^{\star}}}{\left\|\mathbf{h}_{k,k}^{2, \tau_k^{\star}}\right\|_2}$ produces the maximum beamforming gain as  
\[
\left|\left({\bf c}_{k,k}^{2,\tau_{k}^{\star}}\right)^{\sf H}\mathbf{f}_{k}^2\right|^2 =4N,
\]  
while simultaneously eliminating interference toward the adjacent user \( k' \), ensuring  
\[
\left({\bf h}_{k',k}^{2,\tau_{k}^{\star}}\right)^{\sf H}{\bf f}_{k}^2=0.
\]
Fig. \ref{fig:STZF diagram} is a conceptual diagram summarizing the procedure for implementing ST-ZF. As a result, the achievable sum spectral efficiency with the optimal ST-ZF beamforming is given by  
\begin{align}
	R_{\sf sum}^{\sf STZF}=\frac{1}{2}\sum_{k=1}^K\log_2\left(1+\frac{|\beta_{k,k}|^2 2NP}{\sigma^2}\right). \label{eq:SR_3}
\end{align}  
Comparing the sum spectral efficiency of the conventional approach in \eqref{eq:SR_1} and the proposed ST-ZF beamforming in \eqref{eq:SR_2}, we observe that ST-ZF provides a 3 dB SNR gain over the conventional TDMA combined with spatial beamforming. This improvement results from the additional beamforming gain obtained through space-time processing, while still maintaining perfect interference suppression.

\subsection{Discussion on implementation of ST-ZF}

 One potential concern in implementing the proposed ST-ZF is that the Doppler frequency difference between the two associated users may become excessively small, resulting in a retransmission interval, as defined in \eqref{eq:retransmission_interval_of_ST-ZF}, that is excessively large and therefore potentially infeasible. To assess the practical significance of this issue in real-world LEO systems, we conduct an investigation based on the Starlink LEO satellite constellation. Fig. \ref{fig:Satellite_profiles} visualizes the relative motion of Starlink LEO satellites and the corresponding Doppler-related parameters with respect to a typical user located in Seoul. Specifically, it presents (a) the relative velocity of visible satellites, (b) the Doppler frequency offset observed by a selected satellite, and (c) the reciprocal of twice the Doppler frequency offset, which corresponds to the retransmission interval defined in \eqref{eq:retransmission_interval_of_ST-ZF}. The satellite positions and Doppler-related metrics were computed based on Two-Line Element (TLE) data provided by CelesTrak \cite{celestrak2024}, using the Skyfield Python library for high-precision ephemeris propagation \cite{skyfield2024}. 

Fig. \ref{fig:Satellite_profiles}-(a) illustrates that, except for a few outliers, the relative velocities between the typical user and the satellites within the visible region exhibit substantial variation. This variation is sufficient to enable reliable Doppler-based separation of satellites. In contrast, Fig. \ref{fig:Satellite_profiles}-(b) aims to examine the Doppler frequency differences among users associated with the same satellite, from the perspective of the satellite. This observation aligns with the fact that Doppler frequency is strongly dependent on the relative velocity and geometry between the satellite and user. Unlike satellites, which can exhibit large Doppler shifts even when located near each other due to different directions of motion, as shown in Fig. \ref{fig:Satellite_profiles}-(a), users\textemdash who move at extremely low speeds compared to the high-velocity LEO satellites\textemdash tend to experience similar Doppler frequencies when they are geographically close, even if their directions of movement differ. However, due to the extremely high altitude of LEO satellites, each satellite can cover a wide geographical area, thereby ensuring sufficient Doppler separation among users within its coverage footprint. Finally, Fig. \ref{fig:Satellite_profiles}-(c) visualizes the retransmission interval defined in \eqref{eq:retransmission_interval_of_ST-ZF}, which is derived from the Doppler frequency differences illustrated in Fig. \ref{fig:Satellite_profiles}-(b). Considering the system bandwidth, we visualize only the retransmission intervals below 100 $\mu$s, which correspond to the most practically feasible region, and exclude longer intervals as they are unlikely to satisfy the timing requirements of real-time communication systems. Fig. \ref{fig:Satellite_profiles}-(c) indicates that the majority of the coverage area served by the satellite lies within the practically feasible range. Therefore, it can be concluded that, even when the directions of users associated with a satellite are similar, they can still be effectively distinguished in the Doppler domain, and the retransmission interval required for implementing ST-ZF does not pose a significant practical concern.

\section{Space-Time SLNR Beamforming}
\begin{figure}[t!]
  \centering
  \includegraphics[width=0.47\textwidth]{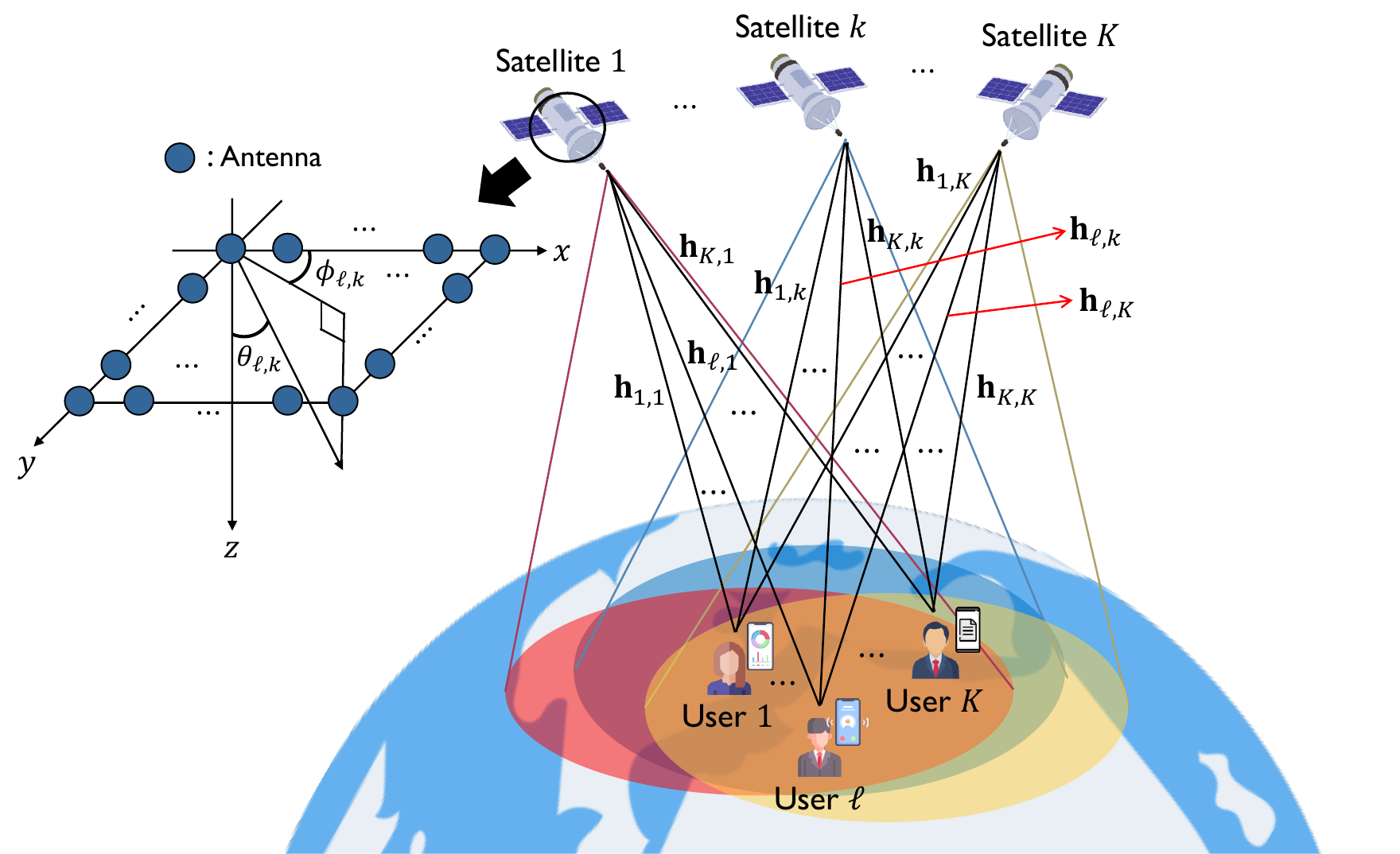}
  \caption{Illustration of the downlink LEO satellite communication system under a fully connected interference channel.}
  \label{fig:Network model} \vspace{-0.2cm}
\end{figure}
 
In this section, we extend the space-time beamforming framework to a more general \( K \)-user interference network as shown in Fig. \ref{fig:Network model}, where each user may experience interference from up to \( K-1 \) satellites, as described in Section II. Additionally, we incorporate a multipath space-time channel model and account for the impact of imperfect channel knowledge at the satellite.  

A major challenge in mitigating inter-beam interference in the \( K \)-user interference network is that each satellite requires at least \( M \) repetitions greater than \( K \) to perfectly eliminate \( K-1 \) interfering signals. However, this leads to a significant loss in sum spectral efficiency. In practical satellite networks, interference suppression is not always necessary for all \( K-1 \) receivers, as treating interference as additional noise can be optimal when the desired link power is sufficiently stronger than the interfering power, as suggested by the conditions in \cite{shang2009new,motahari2009capacity,annapureddy2009gaussian,geng2015optimality,yi2016optimality}. Therefore, an optimal trade-off must be established between interference mitigation and spectral efficiency loss due to signal repetitions.  

To address this, we extend the signal-to-leakage-plus-noise ratio (SLNR) beamforming method. Originally introduced in \cite{sadek2007leakage,sadek2007active}, SLNR beamforming is an interference management technique that maximizes the ratio of the intended signal power at the desired user to the total leaked interference power at unintended users, while also accounting for noise. Unlike ZF and minimum mean square error (MMSE) beamforming, which either completely cancel interference or balance signal power and interference suppression, SLNR beamforming explicitly controls the trade-off between maximizing desired signal power and minimizing interference leakage. This makes it particularly well-suited for interference-limited systems, such as satellite networks where interference levels vary dynamically.

\subsection{Algorithm}
ST-SLNR beamforming requires the joint optimization of three key parameters: the space-time precoding vector \( {\bf f}_k^M \), the repetition interval \( \tau_k \), and the transmission duration \( M \).  

\subsubsection{SLNR Precoding for given $\tau_k$ and $M$}
 The SLNR for satellite $k$ and serving user $k$ over $M$ time slots with time interval $\tau_k$ is defined as 
\begin{align}
	{\sf SLNR}_k\left(M, \tau_k , {\bf f}_k^M \right) &= \frac{\left|\left({\bf h}_{k,k}^{M,\tau_k}\right)^{\sf H}{\bf f}_k^M\right|^2 P}{\sum_{\ell\neq k}^K \left|\left({\bf h}_{\ell,k}^{M,\tau_{k}}\right)^{\sf H}{\bf f}_{k}^M\right|^2 P + M\sigma^2}\nonumber\\
    &=\frac{\left|\left({\bf h}_{k,k}^{M,\tau_k}\right)^{\sf H}{\bf f}_k^M\right|^2 P}{\left\|\left({\bf H}_{k^c,k}^{M,\tau_{k}}\right)^{\sf H}{\bf f}_{k}^M\right\|^2_2 P + M\sigma^2},  \label{eq:SLNR}
\end{align}  
where ${\bf H}_{k^c,k}^{M,\tau_{k}}\in\mathbb{C}^{MN\times K-1}$ is the space-time channel matrix, excluding only ${\bf H}_{k^c,k}^{M,\tau_{k}}$, and is defined as
\begin{equation}
    {\bf H}_{k^c,k}^{M,\tau_{k}}=\begin{bmatrix}
        {\bf h}_{1,k}^{M,\tau_{k}} & \cdots & {\bf h}_{k-1,k}^{M,\tau_{k}} & {\bf h}_{k+1,k}^{M,\tau_{k}} & \cdots & {\bf h}_{K,k}^{M,\tau_{k}} 
    \end{bmatrix}.
\end{equation}
To obtain the SLNR precoding solution, we need to solve the following optimization problem:
\begin{equation}
    \begin{aligned}
       \max_{\{{\bf f}_k^M\}} \quad & {\sf SLNR}_k\left(M, \tau_k , {\bf f}_k^M \right) \\ 
        \text{s.t.} \quad & \|{\bf f}_k^M\|_2^2 = M, \quad k \in [K].
    \end{aligned}
    \label{eq:optimization2}
\end{equation}  
From \cite{sadek2007leakage,sadek2007active}, for given $\tau_k$ and $M$, the optimal precoding solution is given by
\begin{align}
    {\bf \bar f}_k^M =\sqrt{M} \frac{\left( {\bf H}_{k^c,k}^{M,\tau_{k}}\left({\bf H}_{k^c,k}^{M,\tau_{k}}\right)^{\sf H}+\frac{\sigma^2}{P}\mathbf{I}_{MN}\right)^{-1}{\bf h}_{\ell,k}^{M,\tau_k}}{\left\|\left( {\bf H}_{k^c,k}^{M,\tau_{k}}\left({\bf H}_{k^c,k}^{M,\tau_{k}}\right)^{\sf H}+\frac{\sigma^2}{P}\mathbf{I}_{MN}\right)^{-1}{\bf h}_{\ell,k}^{M,\tau_k}\right\|_2}. \label{eq:ST-SLNR beamformer}
\end{align}
This SLNR solution is also known as the virtual uplink MMSE solution.  
   
\subsubsection{Optimization of $\tau_k$}
Invoking the SLNR precoding solution in \eqref{eq:ST-SLNR beamformer} into the definition of SLNR in \eqref{eq:SLNR}, the SLNR for satellite $k$ and serving user $\ell$ boils down to
\begin{align}
	&{\sf SLNR}_k\left(M, \tau_k , 	{\bf \bar f}_k^M \right)\nonumber\\
	&=\left({\bf h}_{\ell,k}^{M,\tau_k}\right)^{\sf H}\left( {\bf H}_{k^c,k}^{M,\tau_{k}}\left({\bf H}_{k^c,k}^{M,\tau_{k}}\right)^{\sf H}+\frac{\sigma^2}{P}\mathbf{I}_{MN}\right)^{-1}{\bf h}_{\ell,k}^{M,\tau_k}.
\end{align}  
The SLNR for satellite $k$ and serving user $\ell$ is only dependent of space-time channel vectors from satellite $k$ to downlink users $\{{\bf H}_{k^c,k}^{M,\tau_{k}}\}_{\ell=1}^K$, implying that each satellite can optimize the time interval $\tau_k$ independently with local CSIT. Accordingly, the optimal $\tau_k$ to maximize SLNR for satellite $k$ and serving user $k$ is obtained by solving the following optimization problem:
\begin{align}
    {\bar \tau}_k &=\argmax_{\tau_k}\left({\bf h}_{k,k}^{M,\tau_k}\right)^{\sf H}\left( {\bf H}_{k^c,k}^{M,\tau_{k}}\left({\bf H}_{k^c,k}^{M,\tau_{k}}\right)^{\sf H}\!\!+\frac{\sigma^2}{P}\mathbf{I}_{MN}\right)^{-1}\!{\bf h}_{k,k}^{M,\tau_k}. \label{eq:optimal time interval}
\end{align}

\begin{algorithm}[t!]
    \caption{ST-SLNR Beamforming}
    Initialization: $M=1,~ R_{k}(0)=0~~\forall k$\;
    Compute ${\bf \bar f}_k^1=\frac{\left( {\bf H}_{k^c,k}^{1,\tau_{k}}\left({\bf H}_{k^c,k}^{1,\tau_{k}}\right)^{\sf H}+\frac{\sigma^2}{P}\mathbf{I}_{N}\right)^{-1}{\bf h}_{k,k}^{1,\tau_k}}{\left\lVert\left( {\bf H}_{k^c,k}^{1,\tau_{k}}\left({\bf H}_{k^c,k}^{1,\tau_{k}}\right)^{\sf H}+\frac{\sigma^2}{P}\mathbf{I}_{N}\right)^{-1}{\bf h}_{k,k}^{1,\tau_k}\right\rVert_2}~~\forall k$\;
    Compute $\sum_{k=1}^{K}R_{k}\left(1,\left\{{\bf \bar f}_k^1\right\}_{k=1}^K\right)$\;
    \While{$\sum_{k=1}^{K}R_{k}\left(M-1,\{\tau_k,\mathbf{f}_k^{M-1}\}_{k=1}^K\right)$ \\ $\qquad\qquad\le\sum_{k=1}^{K}R_{k}\left(M,\{\tau_k,\mathbf{f}_k^M\}_{k=1}^K\right)$}{ 
    $M \leftarrow M+1 $  \;
    \For{ $k = 1, \cdots, K$}{
    ${\bf \bar f}_k^M=\frac{\sqrt{M}\left( {\bf H}_{k^c,k}^{M,\tau_{k}}\left({\bf H}_{k^c,k}^{M,\tau_{k}}\right)^{\sf H}+\frac{\sigma^2}{P}\mathbf{I}_{MN}\right)^{-1}{\bf h}_{k,k}^{M,\tau_k}}{\left\lVert\left( {\bf H}_{k^c,k}^{M,\tau_{k}}\left({\bf H}_{k^c,k}^{M,\tau_{k}}\right)^{\sf H}+\frac{\sigma^2}{P}\mathbf{I}_{MN}\right)^{-1}{\bf h}_{k,k}^{M,\tau_k}\right\rVert_2}$\;
    ${\bar \tau}_k =\underset{\tau_k}{\argmax}\left({\bf h}_{k,k}^{M,\tau_k}\right)^{\sf H}$\\$\qquad\left( {\bf H}_{k^c,k}^{M,\tau_{k}}\left({\bf H}_{k^c,k}^{M,\tau_{k}}\right)^{\sf H}+\frac{\sigma^2}{P}\mathbf{I}_{MN}\right)^{-1}{\bf h}_{k,k}^{M,\tau_k}$\;
    }
    Compute $\sum_{k=1}^{K}R_{k}\left(M,\{\tau_k,\mathbf{f}_k^M\}_{k=1}^K\right)$\;
    }
    \label{algorithm:ST-SLNR}
\end{algorithm}


 \subsubsection{Optimization of $M$}
The final step is to optimize the number of repetitions $M$. The optimal $M$ is obtained by solving the following optimization problem:
\begin{equation}
    \bar{M}=\argmax_{M}\;\sum_{k=1}^K R_{k}\left(M,\left\{\tau_k , {\bf f}_k^M\right\}_{k=1}^K\right), \label{eq: optimal M}
\end{equation}
where formulated based on the sum spectral efficiency that captures the trade-off associated with repeated transmissions. As observed in \eqref{eq:space-time sum rate}, both the SINR and the pre-log term in the achievable spectral efficiency at user $k$ are influenced by $M$. Therefore, the achievable spectral efficiency is determined by which of these two terms is more dominant. When temporal information is first exploited\textemdash namely, when $M$ is a small integer greater than or equal to 2\textemdash the interference suppression effect increases significantly, making the SINR term dominant and leading to an increase in achievable spectral efficiency as $M$ grows. However, beyond a certain point, the achievable spectral efficiency begins to decrease due to the linearly decreasing pre-log term, which outweighs the logarithmic growth of the SINR term. Based on this observation, as shown in Algorithm \ref{algorithm:ST-SLNR}, we monotonically increase $M$ from 1 and compute the sum spectral efficiency. Subsequently, when the sum spectral efficiency begins to decrease after increasing, the computation is terminated, and the value of $M$ that yields the maximum sum spectral efficiency is selected.

\subsection{Extension to Imperfect CSIT Case}
In practical satellite communication systems, obtaining such perfect CSIT challenging due to the rapid movement of satellites and limited CSI feedback link capacity. In this subsection, we consider scenarios with imperfect CSIT to explore the robustness of the proposed approach under such conditions. Let $\hat{\mathbf{h}}_{\ell,k}^{M,\tau_k}$ represent the estimated space-time channel between satellite $k$ and user $\ell$. Then, the space-time channel with imperfect CSIT is modeled as
\begin{equation}
    \hat{\mathbf{h}}_{\ell,k}^{M,\tau_k}={\bf h}_{\ell,k}^{M,\tau_k}-\mathbf{e}_{\ell,k}^M,
\end{equation}
where $\mathbf{e}_{\ell,k}^M$ is the stacked estimation error vector over $M$ time slots as
\begin{equation}
    \mathbf{e}_{\ell,k}^M=\begin{bmatrix}
        \mathbf{e}_{\ell,k}^{\top}[1] & \mathbf{e}_{\ell,k}^{\top}[2] & \cdots & \mathbf{e}_{\ell,k}^{\top}[M]
    \end{bmatrix}^{\top}\in\mathbb{C}^{MN\times 1},
\end{equation}
and it is distributed as independent and identically distributed (IID) complex Gaussian with zero-mean and covariance matrix $\mathbf{\Phi}_{\ell,k}^M=\mathbb{E}\left[\mathbf{e}_{\ell,k}^M\left(\mathbf{e}_{\ell,k}^{M}\right)^{\sf H}\right]\in\mathbb{C}^{MN\times MN}$, i.e., $\mathbf{e}_{\ell,k}^M\sim\mathcal{CN}(\mathbf{0},\mathbf{\Phi}_{\ell,k}^M)$. Furthermore, we assume that all components of $\mathbf{e}_{\ell,k}^M$ are independent, and under these assumptions, $\mathbf{\Phi}_{\ell,k}^M=\sigma_h^2\mathbf{I}_{MN}$ holds for all $k$ and $\ell$. Therefore, due to the estimation error, the coherently combined received signal at user $k$ in \eqref{eq:received signal} can be rewritten as
\begin{align}
y_{k}^M&=\left({\mathbf{h}}_{k,k}^{M,\tau_k}\right)^{\sf H}{\mathbf{f}}_{k}^Ms_k+\sum_{\ell\ne k}^{K}({\mathbf{h}}_{k,\ell}^{M,\tau_k})^{\sf H}{\mathbf{f}}_{\ell}^{M}s_{\ell}+n_{k}^M\nonumber\\
&=\left(\hat{\mathbf{h}}_{k,k}^{M,\tau_k}\right)^{\sf H}{\mathbf{f}}_{k}^M s_k+\sum_{\ell \ne k}^{K}(\hat{\mathbf{h}}_{k,\ell}^{M,\tau_{\ell}})^{\sf H}{\mathbf{f}}_{\ell}^{M}s_q\nonumber\\
&\qquad\qquad\qquad\qquad+\sum_{i=1}^{K}\left(\mathbf{e}_{k,i}^{M}\right)^{\sf H}{\mathbf{f}}_{i}^{M}s_{i}+n_{k}^M.
\end{align}
From this received signal model, the SLNR in \eqref{eq:SLNR} can be rewritten as
\begin{align}
    &{\rm{SLNR}}_{k}(M,\tau_k,{\mathbf{f}}_k^M)\nonumber\\
     &=\frac{\left|\left(\hat{\bf h}_{k,k}^{M,\tau_k}\right)^{\sf H}{\mathbf{f}}_k^M\right|^2P}{\sum_{\ell \ne k}^{K}\left|\left(\hat{\mathbf{h}}_{\ell,k}^{M,\tau_k}\right)^{\sf H}{\mathbf{f}}_k^M\right|^2P+\sum_{i=1}^{K}\left({\mathbf{f}}_i^{M}\right)^{\sf H}\mathbf{\Phi}_{i,k}^M{\mathbf{f}}_i^M P+M\sigma^2}. \label{eq:space-time SLNR imperfect CSIT}
\end{align}
Consequently, under imperfect CSIT, the SLNR precoding solution $\bar{\mathbf{f}}_k^M$ from \eqref{eq:ST-SLNR beamformer} for perfect CSIT is modified as
\begin{align}
    \bar{\mathbf{f}}_k^{M}=\left(\hat{\bf H}_{\ell,k}^{M,\tau_k}\left(\hat{\bf H}_{\ell,k}^{M,\tau_k}\right)^{\sf H}+\left(\sum_{i=1}^K\mathbf{\Phi}_{i,k}^M+\frac{\sigma^2}{P}\right)\mathbf{I}_{MN}\right)^{-1}\hat{\bf h}_{k,k}^{M,\tau_k},
\end{align}
where
\begin{equation}
    \hat{\bf H}_{k^c,k}^{M,\tau_{k}}=\begin{bmatrix}
        \hat{\bf h}_{1,k}^{M,\tau_{k}} & \cdots & \hat{\bf h}_{k-1,k}^{M,\tau_{k}} & \hat{\bf h}_{k+1,k}^{M,\tau_{k}} & \cdots & \hat{\bf h}_{K,k}^{M,\tau_{k}} 
    \end{bmatrix}.
\end{equation}
Subsequently, as with the perfect CSIT case, determining $\bar{\tau}_k$ and $\bar M$ under imperfect CSIT conditions using the SLNR precoding solution $\bar{\mathbf{f}}_k^{M}$ is based on the criteria outlined in \eqref{eq:optimal time interval} and \eqref{eq: optimal M}.

\section{Simulation Results} \label{sec:Simulation}
In this section, we evaluate the performance of the proposed space-time beamforming technique through simulations. We first describe the simulation setup used in this study, detailing the key system parameters and assumptions. Then, we present the simulation results in two different network scenarios, highlighting the performance improvements over traditional beamforming approaches.

\subsection{Simulation Setup}

\subsubsection{Channel Model}
The attenuation constant is determined by the path-loss due to the satellite-user distance, the channel fading effects, including both shadowing and small-scale fading, and the tap gain that accounts for path-specific attenuation in a multi-path environment.

We adopt a classical path-loss model that depends on the satellite-user distance. The path-loss for the \( i \)th path between satellite \( k \) and user \( \ell \), operating at carrier frequency \( f_c \), is given by  
\begin{equation}
    D_{\ell,k,i}=\left(\frac{c}{4\pi f_c d_{\ell,k,i}}\right)^{\alpha},
\end{equation}  
where \( c \) is the speed of light, \( d_{\ell,k,i} \) is the distance of the \( i \)th path, and \( \alpha \) is the path-loss exponent. To model shadowing and small-scale fading, we use the shadowed-Rician distribution, a well-established model for land mobile satellite (LMS) channels \cite{abdi2003new}. This model represents the LoS component using a Nakagami distribution and the scatter component using a Rayleigh distribution. It effectively captures channel conditions ranging from unobstructed LoS to environments with complete or partial blockage.  

Following \cite{abdi2003new}, we consider the average shadowing scenario, one of three representative cases (frequent heavy shadowing, average shadowing, and infrequent light shadowing). By integrating path-loss, fading effects, and tap gain, the attenuation constant for the \( i \)th path between satellite \( k \) and user \( \ell \) is modeled as  
\begin{equation}
    \beta_{\ell,k,i}=\delta^{i-1}\sqrt{D_{\ell,k,i}H_{\ell,k,i}},
\end{equation}  
where \( H_{\ell,k,i} \) represents the shadowed-Rician fading power for the \( i \)th path. The parameter \( \delta < 1 \) denotes the tap gain, assuming an exponentially decaying path structure, where the \( i \)th path corresponds to the \( i \)th tap reaching the user for \( i \in \{1,2,\dots,L_{\ell,k}\} \).

\subsubsection{Network Model and Assumptions}
 
We consider satellites equipped with UPAs (\( N_x = N_y = 8 \)) operating at 1.9925 GHz with a 5 MHz bandwidth, positioned at a 530 km altitude, based on SpaceX's FCC specifications for direct-to-cellular systems \cite{FCC2023direct}. The speed of light, noise spectral density, and path-loss exponent are set to \( 3\times10^5 \) km/s, -174 dBm/Hz, and 2, respectively. Users are assumed to be near \( (0,0,R_{\sf E}) \), where \( R_{\sf E} = 6371 \) km denotes the Earth's radius. The reference satellite is located at an azimuth of \( 10^\circ \) and a zenith of \( 5^\circ \) relative to the Earth's center, with other satellites distributed nearby.  

To model dynamic satellite-user interactions, their positions are randomly distributed around reference points, with the user-satellite direction uniformly varying within \( \pm 1^\circ \) in azimuth and zenith angles. The LoS path follows this direction (\( i=1 \)), while other multipath components deviate randomly within \( \pm 1^\circ \). Doppler shifts, influenced by relative velocity and altitude, are uniformly distributed between -50 kHz and 50 kHz \cite{nguyen2021tcp,kabir2023precise}, and assumed identical across all channel paths. Each satellite-user channel comprises \( L = 3 \) multipath components (\( L_{\ell,k} = L \)). We set \( \delta = 0.5 \) and assume the channel estimation error variance \( \sigma_h^2 \) equals the noise variance \( \sigma^2 \) to evaluate robustness under imperfect CSIT. For performance comparison, we benchmark the proposed space-time beamforming against conventional MRT, ZF, SLNR-based spatial beamforming, and TDMA.

\subsection{Numerical Results for Partially Connected Networks}
\begin{figure}[t!]
  \centering
  \includegraphics[width=0.47\textwidth]{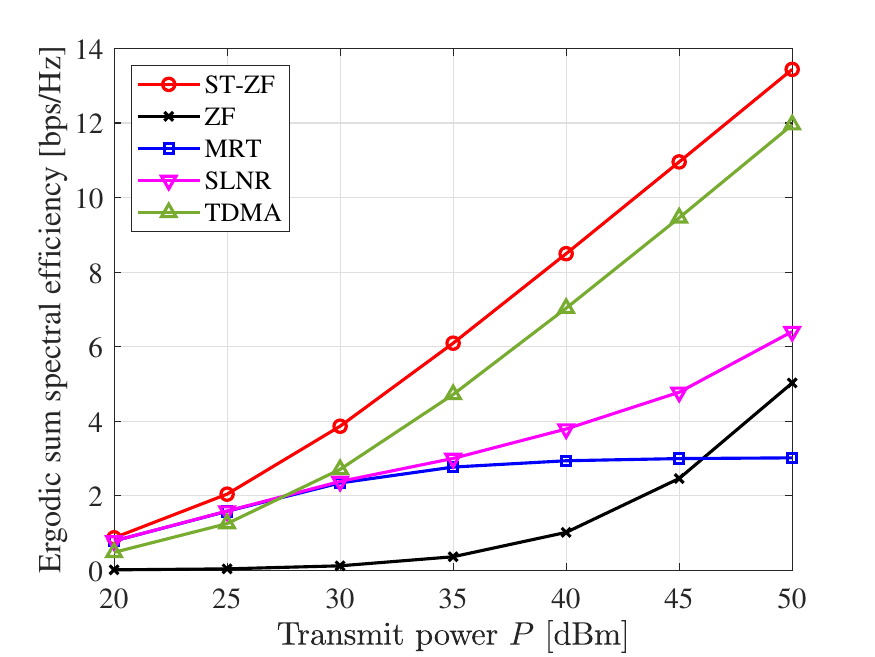}
  \caption{Ergodic sum spectral efficiency as a function of $P$ in a partially connected interference network ($K=3$).}
  \label{fig:sim-Power partial multi} \vspace{-0.2cm}
\end{figure}

{\bf Effect of Beamforming Methods}: Fig. \ref{fig:sim-Power partial multi} illustrates the ergodic sum spectral efficiency for a partially connected interference network as a function of $P$ when $K=3$. First, with respect to the effectiveness of beamforming strategies, ST-ZF outperforms conventional spatial beamforming methods as well as TDMA. This is because, in dense LEO satellite networks where satellites serve users that are co-located, the users appear in similar directions from the perspective of the satellites. As a result, relying solely on spatial information for beamforming is insufficient for effective interference management. Therefore, in such environments, TDMA\textemdash considered information-theoretically optimal and simple to implement\textemdash is commonly employed. However, as demonstrated in Fig. \ref{fig:sim-Power partial multi} and compared in \eqref{eq:SR_1} and \eqref{eq:SR_3}, the proposed ST-ZF achieves a 3 dB SNR gain over TDMA. This arises from the fundamentally different inherent characteristics of the two beamforming approaches. Although both methods utilize two time slots, in TDMA, users receive signals from only one time slot to manage interference. In contrast, ST-ZF enables users to receive signals across both time slots while perfectly suppressing interference, thereby simultaneously achieving the benefits of ZF and repetition coding.

{\bf Effect of $P$}: As observed in Fig. \ref{fig:sim-Power partial multi}, spatial beamforming based on MRT converges to a specific value as the transmit power increases, due to the concurrent increase in interference. Based on this observation, in the high SNR regime, spatial beamforming based on MRT naturally exhibits the poorest performance. ZF and SLNR-based spatial beamforming perform better than MRT in the high SNR regime but still exhibit inferior performance compared to space-time beamforming and TDMA due to their low spatial resolution. In partially connected interference networks, the proposed ST-ZF can also serve all users with two repeated signal transmissions. This indicates that ST-ZF, like TDMA, can achieve the information-theoretic DoF limit of $\frac{K}{2}$ in a partially connected interference network. Therefore, the proposed ST-ZF is optimal from a DoF perspective. Consequently, in the high SNR regime, the slope of the sum spectral efficiency with respect to transmit power is the same for ST-ZF and TDMA. Combined with the effectiveness of the beamforming strategies, this confirms that the proposed ST-ZF is DoF-optimal and achieves a 3 dB gain over TDMA, thereby demonstrating its superiority over conventional methods.

\begin{figure}[t!]
  \centering
  \includegraphics[width=0.47\textwidth]{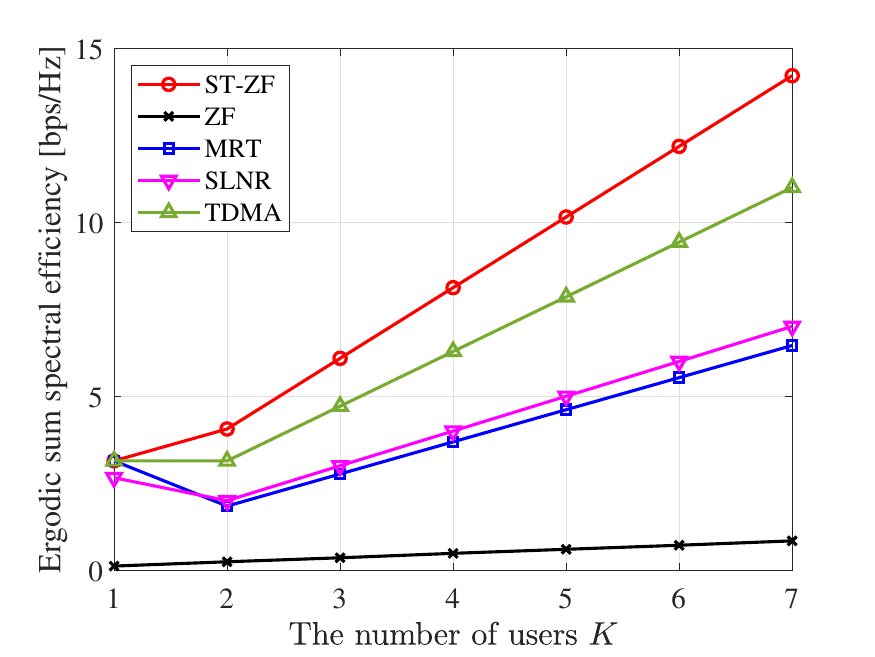}
  \caption{Ergodic sum spectral efficiency as a function of $K$ in a partially connected interference network ($P=35$ dBm).}
  \label{fig:sim-number of users partial multi} \vspace{-0.2cm}
\end{figure}

{\bf Effect of $K$}: Next, Fig. \ref{fig:sim-number of users partial multi} illustrates the ergodic sum spectral efficiency for a partially connected interference network as a function of $K$ when $P=35$ dBm. In contrast to Fig. \ref{fig:sim-Power partial multi}, where the pre-log term served as the slope in the sum spectral efficiency with respect to transmit power $P$, Fig. \ref{fig:sim-number of users partial multi} shows that the log term containing the SINR becomes the slope of the sum spectral efficiency with respect to the number of users $K$. Therefore, with a 3 dB SNR gain per user, ST-ZF exhibits a steeper slope than TDMA, leading to a more rapid increase in sum spectral efficiency as the number of users increases.

\subsection{Numerical Results for Fully Connected Networks}
\begin{figure}[t!]
  \centering
  \includegraphics[width=0.47\textwidth]{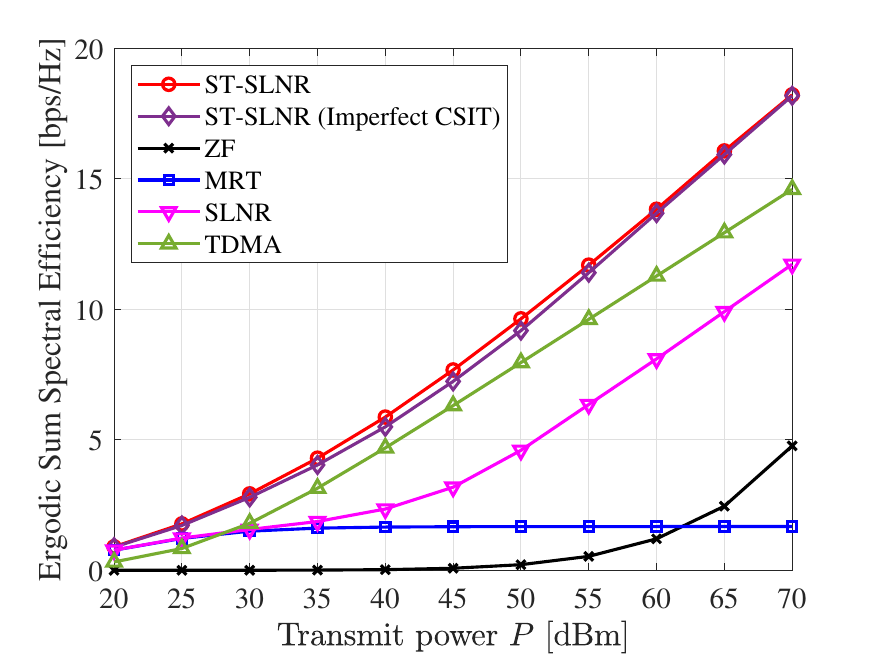}
  \caption{Ergodic sum spectral efficiency as a function of $P$ in a fully connected interference network ($M=3$ and $K=4$).}
  \label{fig:sim-Txpower multi} \vspace{-0.2cm}
\end{figure}

{\bf Effect of Beamforming Methods}: Fig. \ref{fig:sim-Txpower multi} illustrates the ergodic sum spectral efficiency for a fully connected interference network as a function of $P$ when $M=3$ and $K=4$. According to the ST-SLNR algorithm described in Section \uppercase\expandafter{\romannumeral4}-A, when $K=4$, the value of $M$ is set to 3, which yields the highest sum spectral efficiency. Since a fully connected interference network is a generalized model of a partially connected interference network, in dense LEO satellite networks, it not only inherits the difficulty of distinguishing co-located users based on direction\textemdash as in the partially connected case\textemdash but also experience even greater levels of interference. Therefore, as in the case of the partially connected interference network, conventional spatial beamforming based solely on directional information exhibits very poor performance. In this network model, even when multiple time slots are employed, TDMA\textemdash which can completely avoid interference\textemdash is more effective than spatial beamforming. However, the proposed ST-SLNR, which jointly leverages both spatial and temporal information, constitutes a fundamentally novel beamforming approach and outperforms TDMA. Moreover, the proposed ST-SLNR demonstrates robustness even under imperfect CSIT conditions. This is because, even in the presence of channel errors, the satellite can optimize the repetition time interval accordingly.

{\bf Effect of $P$}: In the fully connected interference network, the spatial beamforming methods exhibits the same performance trend as observed in Fig. \ref{fig:sim-Power partial multi} for the partially connected interference network. Therefore, in this case as well, TDMA outperforms the spatial beamforming methods. However, in a fully connected interference network, it is not feasible for TDMA to serve all users within two time slots as it does in the partially connected interference network. In this network, TDMA must serve only one user per time slot, resulting in a pre-log term of 1. In contrast, ST-SLNR in Fig. \ref{fig:sim-Txpower multi} can simultaneously serve four users using three time slots, resulting in a pre-log term of $\frac{4}{3}>1$. Consequently, in the high SNR regime, the proposed ST-SLNR exhibits a steeper slope than TDMA, leading to a more rapid increase in sum spectral efficiency as the transmit power increases.

\begin{figure}[t!]
  \centering
  \includegraphics[width=0.47\textwidth]{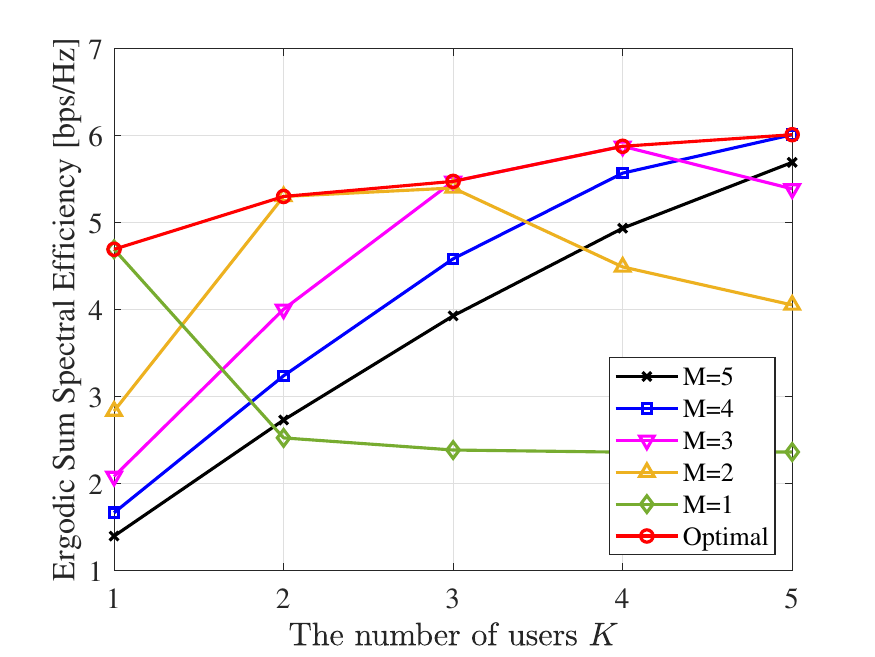}
  \caption{Ergodic sum spectral efficiency of ST-SLNR as a function of $K$ and $M$ in a fully connected interference network ($P=40$ dBm).}
  \label{fig:sim-number of users multipath M} \vspace{-0.2cm}
\end{figure}

{\bf Effect of $M$}: Fig. \ref{fig:sim-number of users multipath M} illustrates the ergodic sum spectral efficiency of ST-SLNR for a fully connected interference network as a function of $K$ and $M$ when $P=40$ dBm. Fig. \ref{fig:sim-number of users multipath M} highlights the trade-off inherent in space-time beamforming between the enhancement in interference suppression achieved by leveraging temporal information and the inevitable reduction in spectral efficiency caused by transmitting the same data multiple times. As explained earlier, when $K\ge 2$, where interference begins to arise, setting $M$ to a sufficiently small value greater than or equal to 2 allows users to be distinguished through additional signatures derived from temporal information, resulting in a rapid enhancement in interference suppression. Consequently, the gain in SINR becomes dominant, leading to an increase in sum spectral efficiency. However, beyond a certain point, unlike the logarithmic growth of the SINR, the pre-log term decreases linearly, eventually causing the sum spectral efficiency to decline. For example, when $K=4$, the sum spectral efficiency increases up to $M=3$; however, starting from $M=4$, the pre-log term becomes more dominant than the SINR gain, resulting in a continued decline in sum spectral efficiency. Based on this result, $M$ is set to 3 when $K=4$ in Fig. \ref{fig:sim-Txpower multi}.

\begin{figure}[t!]
  \centering
  \includegraphics[width=0.47\textwidth]{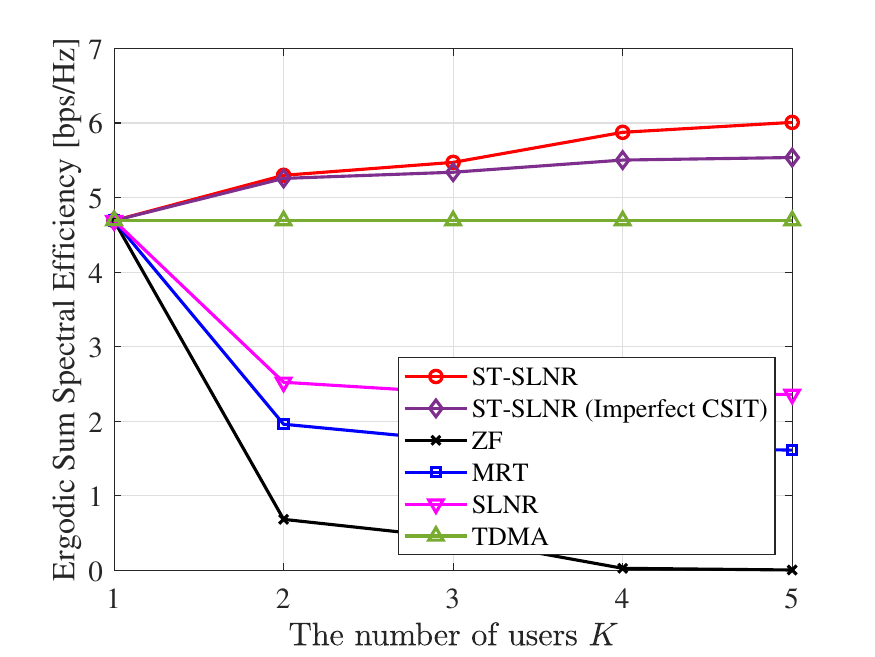}
  \caption{Ergodic sum spectral efficiency as a function of $K$ in a fully connected interference network ($P=40$ dBm).}
  \label{fig:sim-number of users multipath} \vspace{-0.2cm}
\end{figure}

{\bf Effect of $K$}: Finally, Fig. \ref{fig:sim-number of users multipath} shows the ergodic sum spectral efficiency for a fully connected interference network as a function of $K$ when $P=40$ dBm. A notable distinction from the partially connected interference network in Fig. \ref{fig:sim-number of users partial multi} is that, in the fully connected interference network, the pre-log term of TDMA remains fixed at 1 regardless of the number of users. As a result, the sum spectral efficiency does not increase with the number of users. In contrast, ST-SLNR not only benefits from increased SINR due to interference suppression, but also exhibits pre-log terms greater than 1 for certain values of $K$. Consequently, the sum spectral efficiency continues to increase as the number of users grows. Additionally, since the pre-log term of ST-SLNR is always greater than or equal to 1, it is never smaller than that of TDMA. Combined with its interference suppression capability, ST-SLNR consistently outperforms TDMA. Furthermore, consistent with the observations in Fig. \ref{fig:sim-Txpower multi}, Fig. \ref{fig:sim-number of users multipath} also confirms that ST-SLNR is robust to channel estimation errors under imperfect CSIT conditions.

\section{Conclusion} \label{sec:conclusion}
In this paper, we proposed a space-time beamforming technique to mitigate mutual inter-beam interference that occurs when multiple satellites simultaneously serve co-located users. Users co-located within the spatial resolution cannot be spatially distinguished, rendering conventional spatial beamforming techniques ineffective for interference mitigation. Consequently, a TDMA scheme based on a cake-cutting approach, which orthogonalizes channel access in the time domain, is practically employed. The proposed space-time beamforming technique effectively manages inter-beam interference in such environments, achieving superior performance compared to both conventional spatial beamforming and TDMA in fully connected interference channels as well as their special case, partially connected interference channels. We have validated this performance improvement through simulation results. This superior inter-beam interference management capability originates from the space-time transmission strategy. This strategy, which involves repeatedly transmitting signals and co-processing them, allows for the synthesis of a virtual array larger than the physically limited aperture size of antennas deployable on satellites, enabling the formation of narrower beams compared to traditional methods. Furthermore, the Doppler characteristics naturally generated by repeated transmissions serve as temporal signatures, providing an additional dimension to distinguish co-located users. However, the repeated transmission of the same signal, while enabling the acquisition of temporal signatures from Doppler frequencies, comes at the cost of reduced spectral efficiency. In this paper, we introduce the optimal number of repeated transmissions and time intervals that best balance this trade-off for both fully connected and partially connected interference channels.


\bibliographystyle{IEEEtran}
\bibliography{refs_all}


 





\end{document}